\documentclass[twocolumn,aps,prb,showpacs,superscriptaddress,floatfix]{revtex4}

\usepackage{graphicx}
\usepackage{dcolumn}
\usepackage{bm}
\begin{document}
\preprint{}

\title{Dilute Magnetism and Spin-Orbital Percolation Effects in Sr$_2$Ir$_{1-x}$Rh$_x$O$_4$}

\author{J.P. Clancy}
\affiliation{Department of Physics, University of Toronto, Toronto, Ontario, M5S 1A7, Canada}

\author{A. Lupascu}
\affiliation{Department of Physics, University of Toronto, Toronto, Ontario, M5S 1A7, Canada}

\author{H. Gretarsson}
\affiliation{Department of Physics, University of Toronto, Toronto, Ontario, M5S 1A7, Canada}

\author{Z. Islam}
\affiliation{Advanced Photon Source, Argonne National Laboratory, Argonne, Illinois 60439, USA}

\author{Y.F. Hu}
\affiliation{Canadian Light Source, Saskatoon, Saskatchewan, S7N 0X4, Canada}

\author{D. Casa}
\affiliation{Advanced Photon Source, Argonne National Laboratory, Argonne, Illinois 60439, USA}

\author{C.S. Nelson}
\affiliation{National Synchrotron Light Source, Brookhaven National Laboratory, Upton, New York, 11973, USA}

\author{S.C. LaMarra}
\affiliation{National Synchrotron Light Source, Brookhaven National Laboratory, Upton, New York, 11973, USA}

\author{G. Cao}
\affiliation{Department of Physics and Astronomy, University of Kentucky, Lexington, Kentucky 40506, USA}

\author{Young-June Kim}
\affiliation{Department of Physics, University of Toronto, Toronto, Ontario, M5S 1A7, Canada}

\begin{abstract}
We have used a combination of resonant magnetic x-ray scattering (RMXS) and x-ray absorption spectroscopy (XAS) to investigate the properties of the doped spin-orbital Mott insulator Sr$_2$Ir$_{1-x}$Rh$_x$O$_4$ (0.07 $\le$ x $\le$ 0.70). We show that Sr$_2$Ir$_{1-x}$Rh$_x$O$_4$ represents a unique model system for the study of dilute magnetism in the presence of strong spin-orbit coupling, and provide evidence of a doping-induced change in magnetic structure and a suppression of magnetic order at x$_c$ $\sim$ 0.17. We demonstrate that Rh-doping introduces Rh$^{3+}$/Ir$^{5+}$ ions which effectively hole-dope this material. We propose that the magnetic phase diagram for this material can be understood in terms of a novel spin-orbital percolation picture.  
\end{abstract}


\maketitle

\section{Introduction}

The physics of iridium-based transition metal oxides has sparked significant interest due to the potential for exotic electronic and magnetic ground states driven by strong spin-orbit coupling (SOC).  Due to the large atomic mass and broad electronic wavefunctions associated with 5d iridium, these materials tend to display strong relativistic SOC and crystal electric field (CEF) effects, but relatively weak electronic correlations (U).  As a result, the properties of the 5d iridates are often dramatically different from those of their lighter 3d counterparts.  The layered perovskite Sr$_2$IrO$_4$ has attracted particular attention as the first experimental realization of a $J_{eff}$ = 1/2 spin-orbital Mott insulator \cite{Kim_PRL_2008, Kim_Science_2009}.  The magnetism in this compound arises from Ir$^{4+}$ ions with a 5d$^5$ electronic configuration.  However, unlike conventional S = 1/2 magnetic moments, the $J_{eff}$ = 1/2 moments of Sr$_2$IrO$_4$ possess mixed spin and orbital character, with magnetic exchange interactions that are strongly bond and lattice-dependent.  For the bond geometry of Sr$_2$IrO$_4$, these interactions can be described by an effectively isotropic Heisenberg Hamiltonian \cite{Jackeli_PRL_2009,Kim_PRL_2012}. 

Sr$_2$IrO$_4$ has a tetragonal crystal structure (space group {\it I4$_1$/acd}, {\it a} = 5.499 {\AA}, {\it c} = 25.79 {\AA}) which consists of stacked layers of corner-sharing IrO$_6$ octahedra \cite{Crawford_PRB_1994,Huang_JSSC_1994}.  This structure is a variant of the K$_2$NiF$_4$ structure shared by La$_{2-x}$(Ba,Sr)$_x$CuO$_4$ and Sr$_2$RuO$_4$, differing only by a staggered $\sim$ 11$^{\circ}$ rotation of IrO$_6$ octahedra about the {\it c}-axis.  The structural and magnetic similarities between these compounds have led to natural associations with superconductivity, and recent theoretical proposals \cite{Wang_PRL_2011,Watanabe_PRL_2013} have spurred renewed interest in the properties of doped Sr$_2$IrO$_4$.  Although many forms of electron, hole, and isoelectronic doping have been experimentally tested to date \cite{Qi_PRB_2012,Lee_PRB_2012,Klein_JPCM_2008,Klein_JEM_2009,Cao_APS_2013,Calder_PRB_2012,Carter_PRB_1995,Cava_PRB_1994,Gatimu_JSSC_2012,Ge_PRB_2011,Korneta_PRB_2010,Shimura_PRB_1995}, there is still much to learn about the impact of doping on the spin-orbital Mott insulating ground state.

Sr$_2$Ir$_{1-x}$Rh$_{x}$O$_4$ represents an ideal candidate for experimental doping studies.  Rh is situated directly above Ir in the periodic table, and Sr$_2$RhO$_4$ is a paramagnetic metal which is isostructural to Sr$_2$IrO$_4$ (with slightly reduced lattice parameters and an octahedral rotation of $\sim$ 9.7$^{\circ}$) \cite{Perry_NJP_2006,Martins_PRL_2011,Itoh_JSSC_1995}.  Bulk characterization measurements on Sr$_2$Ir$_{1-x}$Rh$_x$O$_4$ have revealed a rich phase diagram with multiple electronic and magnetic transitions \cite{Qi_PRB_2012}.  At low concentrations (x $\le$ 0.16), Sr$_2$Ir$_{1-x}$Rh$_x$O$_4$ is an antiferromagnetic insulator, while at higher dopings it becomes a paramagnetic metal/semiconductor (0.16 $\le$ x $\le$ 0.24), a frustrated magnetic insulator (0.24 $\le$ x $\le$ 0.85), and a paramagnetic correlated metal (x $\ge$ 0.85).  In the simplest scenario, one expects Rh-doping to result in an isoelectronic substitution of Ir$^{4+}$ (5d$^5$) for Rh$^{4+}$ (4d$^5$).  Such a substitution would tune the SOC of the system from the strong 5d regime to the moderate 4d regime, but leave the band filling unaffected \cite{Qi_PRB_2012,Lee_PRB_2012}.  However, it has also been proposed that the dopant ions may adopt a Rh$^{3+}$ (4d$^6$) oxidation state, creating nearby Ir$^{5+}$ (5d$^4$) ions in order to preserve charge neutrality \cite{Klein_JPCM_2008,Klein_JEM_2009,Cao_APS_2013}.  Such a substitution would not only tune SOC, but would also alter the band filling via hole-doping.  A comparison of these two mechanisms is provided in Fig. 1(a).  Rh$^{4+}$ and Rh$^{3+}$ substitution will also have very different effects on the magnetism of Sr$_2$IrO$_4$, with Rh$^{4+}$ doping resulting in an exchange of effective S = 1/2 moments, and Rh$^{3+}$ doping introducing pairs of non-magnetic vacancies (Rh$^{3+}$ and Ir$^{5+}$ are both non-magnetic due to fully filled t$_{2g}$ [Rh] and $J_{eff}$ = 3/2 [Ir] electronic configurations).  

\begin{figure}
\includegraphics{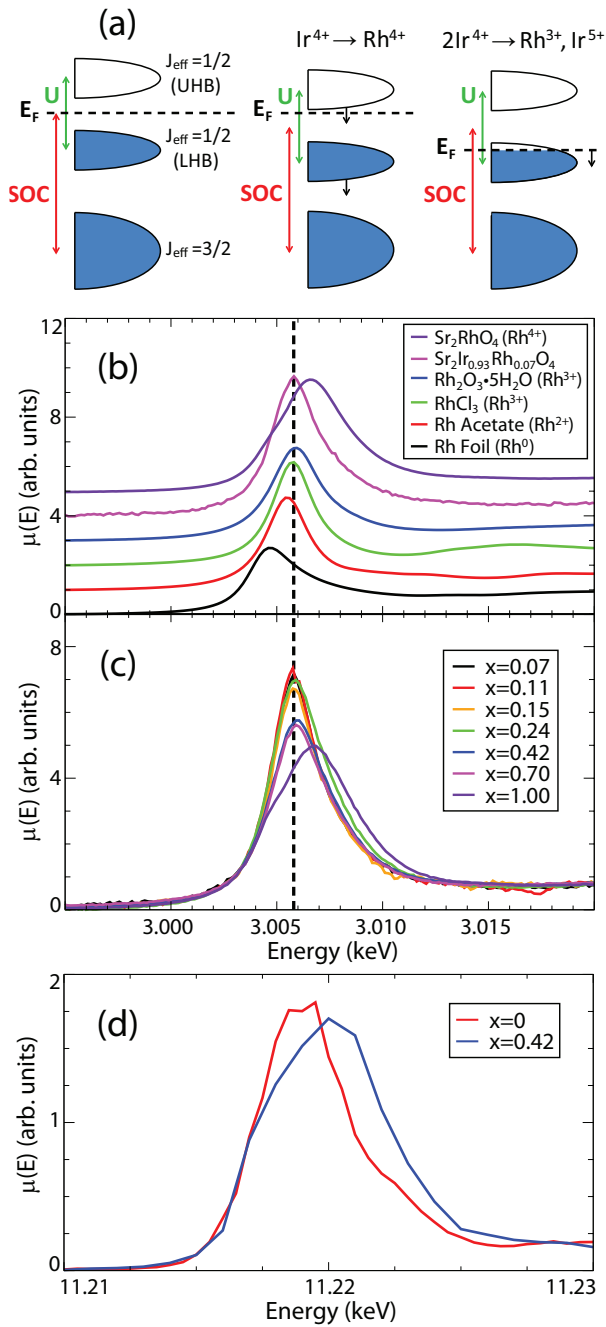} 
\caption{(Color Online) (a) Potential effects of Rh-doping on the electronic band structure of Sr$_2$IrO$_4$.  Isoelectronic substitution of Rh$^{4+}$ for Ir$^{4+}$ is expected to tune the strength of the SOC, while substitution of Rh$^{3+}$ is expected to reduce SOC and effectively hole-dope the system.  (b) X-ray absorption spectra collected at the Rh L$_3$-edge (2p$_{3/2}$ $\rightarrow$ 4d) for a series of Rh-based reference samples.  The position of the white-line peak in Sr$_2$Ir$_{1-x}$Rh$_x$O$_4$ is consistent with a Rh$^{3+}$ oxidation state.  (c) Doping dependence of Rh L$_3$-edge absorption spectra for Sr$_2$Ir$_{1-x}$Rh$_x$O$_4$.  The position of the white-line feature remains unchanged for 0.07 $\le$ x $\le$ 0.70.  (d) X-ray absorption spectra collected at the Ir L$_3$-edge (2p$_{3/2}$ $\rightarrow$ 5d) for samples with x = 0 and x = 0.42.  The positive chemical shift and the broadening of the white-line feature are consistent with a mixed population of Ir$^{4+}$ and Ir$^{5+}$ ions introduced by doping.}
\end{figure}

In this article, we present complementary resonant magnetic x-ray scattering (RMXS) and x-ray absorption spectroscopy (XAS) measurements on single crystal samples of Sr$_2$Ir$_{1-x}$Rh$_x$O$_4$ (0.07 $\le$ x $\le$ 0.70).  Our results clearly demonstrate that Sr$_2$Ir$_{1-x}$Rh$_x$O$_4$ must be considered as a hole-doped and magnetically diluted system.  We show that Rh-doping results in a change of magnetic structure, a rapid decrease in magnetic transition temperatures, and a suppression of magnetic order at x$_c$ $\sim$ 0.17.  In contrast to diluted La$_2$CuO$_4$, we show that the magnetic phase diagram of Sr$_2$Ir$_{1-x}$Rh$_x$O$_4$ cannot be described by a conventional spin-only percolation picture.  We propose that this discrepancy may reflect the importance of both spin and orbital percolation effects, which arise due to the strong SOC inherent to this system.

\section{Experimental Details}

Single crystal samples of Sr$_2$Ir$_{1-x}$Rh$_x$O$_4$ (0 $\le$ x $\le$ 1.0) were prepared using self-flux techniques, as described elsewhere \cite{Cao_PRB_1998,Qi_PRB_2012}.  The samples used in this experiment had typical dimensions of $\sim$ 2.0 $\times$ 1.0 $\times$ 0.1 mm$^3$.  Detailed magnetization, resistivity, and specific heat measurements on these samples have previously been reported by Qi et al \cite{Qi_PRB_2012}.  The Rh content of each sample was determined by energy dispersive x-ray (EDX) spectroscopy using a combined Hitachi/Oxford SwiftED 3000 unit.  Crystal quality was assessed by x-ray rocking scans, which revealed sample mosaicities of $\sim$ 0.01$^{\circ}$ to 0.15$^{\circ}$ full-width at half-maximum (FWHM).  In particular, the three samples which lie within the magnetically-ordered region of the phase diagram (x = 0.07, 0.11, and 0.15) all displayed FWHM of 0.02$^{\circ}$ or better. 

X-ray absorption spectroscopy measurements were performed using the Soft X-ray Microcharacterization Beamline (SXRMB) at the Canadian Light Source (CLS) and Beamline 9-ID-B at the Advanced Photon Source (APS) at Argonne National Laboratory.  Measurements on SXRMB were carried out at the Rh L$_3$ (2p$_{3/2}$ $\rightarrow$ 4d) and L$_2$ (2p$_{1/2}$ $\rightarrow$ 4d) absorption edges, which occur at energies of 3.004 keV and 3.146 keV respectively.  Data was collected using Total Electron Yield (TEY) and Fluorescence Yield (FY) detection modes, and energy calibration was verified by a comparison of Ar K-edge features observed at E = 3.206 keV.  Measurements on 9-ID-B were carried out at the Ir L$_3$ absorption edge (2p$_{3/2}$ $\rightarrow$ 5d, E = 11.215 keV), using Partial Fluorescence Yield (PFY) detection mode.  PFY-XAS is a form of resonant x-ray emission spectroscopy, which involves tuning the incident energy to the Ir L$_3$-edge, and monitoring the intensity of the Ir L$\alpha_2$ emission line (3d$_{3/2}$ $\rightarrow$ 2p$_{3/2}$, E = 9.099 keV) as a function of energy.  By suppressing the spectral broadening due to 2p core-hole lifetime effects, PFY-XAS can provide a significant improvement in experimental energy resolution\cite{Hamalainen_PRL_1991,deGroot_PRB_2002}.  These measurements were carried out using a double-bounce Si-(1,1,1) primary monochromator, a channel-cut Si-(3,3,3) secondary monochromator, and a spherical (1m radius) diced Ge-(3,3,7) analyzer crystal to obtain an instrumental energy resolution of 225 meV (FWHM).  Measurements were collected using horizontal scattering geometry, with a scattering angle close to 2$\theta$ = 90$^{\circ}$.

Resonant magnetic x-ray scattering measurements were performed using Beamline 6-ID-B at the APS.  Measurements were carried out at the Ir L$_3$ (2p$_{3/2}$ $\rightarrow$ 5d) and L$_2$ (2p$_{1/2}$ $\rightarrow$ 5d) absorption edges, which occur at energies of 11.215 keV and 12.824 keV respectively.  Samples were mounted in a closed-cycle cryostat with a base temperature of T = 6 K.  Measurements were performed in vertical scattering geometry, with the polarization of the incident beam perpendicular to the scattering plane defined by {\bf k$_i$} and {\bf k$_f$} (i.e. a $\sigma$-polarized beam).  The polarization of the scattered beam was analyzed using the (0,0,8) and (0,0,10) reflections from a pyrolytic graphite (PG) analyzer crystal.  These reflections correspond to analyzer scattering angles of $2\theta_p$ = 82.33$^{\circ}$ at the Ir L$_3$-edge and $2\theta_p$ = 92.04$^{\circ}$ at the Ir L$_2$-edge, respectively.  In this configuration, the scattering term with a rotated polarization vector (i.e. $\sigma$-$\pi$) is magnetic in origin, while the term with an unrotated polarization vector (i.e. $\sigma$-$\sigma$) is due to charge scattering.  The intensity of the magnetic scattering contribution is proportional to ({\bf k$_f$} $\cdot$ {\bf M})$^2$ [Ref. 27].

High-resolution non-resonant x-ray diffraction measurements were performed using Beamline X21 at the National Synchrotron Light Source (NSLS) at Brookhaven National Laboratory.  Measurements were carried out in vertical scattering geometry, using x-rays with an incident energy of 11.000 keV.  A Ge-(1,1,1) analyzer was used to improve angular resolution and reduce experimental background.

\section{Experimental Results}

\subsection{Ionic Composition of S\lowercase{r}$_2$I\lowercase{r}$_{1-x}$R\lowercase{h}$_{x}$O$_4$}

To investigate the role of the Rh dopant ions in Sr$_2$Ir$_{1-x}$Rh$_x$O$_4$ we performed x-ray absorption spectroscopy (XAS) measurements at the Rh L$_3$-edge.  The position of the sharp ``white-line'' peak at the absorption edge is very sensitive to oxidation state, and displays a chemical shift which is proportional to the ionic charge.  Fig. 1(b) shows representative x-ray absorption spectra for a series of Rh-based reference samples, with oxidation states ranging from 0 to 4+.  The white-line peak for Sr$_2$Ir$_{0.93}$Rh$_{0.07}$O$_4$ clearly coincides with the Rh$^{3+}$ reference samples, and is shifted by $\sim$ -1.4 eV with respect to Rh$^{4+}$.  The doping dependence of the absorption spectra (Fig. 1(c)) indicates that the position of the white-line peak remains fixed for x = 0.07 to x = 0.70.  

By performing similar XAS measurements at the Ir L$_3$-edge (Fig. 1(d)) we can also characterize the doping dependence of the Ir oxidation state.  These measurements reveal a broadening of the Ir L$_3$-edge white-line peak and a positive shift in spectral weight with increasing Rh concentration.  Both of these features are consistent with a mixed population of Ir$^{4+}$ and Ir$^{5+}$ ions introduced by doping.  It should be noted that the doping dependence of the Ir L$_3$-edge absorption spectra is difficult to observe with conventional XAS methods due to the effect of core-hole lifetime broadening, which is more than twice as large for Ir (5.3 eV) as it is for Rh (2.1 eV)\cite{Fuggle_and_Inglesfield}.  It is only by utilizing the PFY-XAS technique, which suppresses such core-hole lifetime effects, that we can resolve these features in the present study. 

The combination of Rh and Ir XAS results allow us to draw four main conlusions: (1) The Rh dopant ions in Sr$_2$Ir$_{1-x}$Rh$_x$O$_4$ adopt a 3+ rather than 4+ oxidation state.  (2) This oxidation state persists across almost the entire Rh-doped phase diagram.  (3) The electronic effect of Rh-doping is to tune band-filling via hole-doping.  (4) The magnetic effect of Rh-doping is to introduce quenched non-magnetic vacancies (2 per dopant ion).

We must emphasize that while the Rh$^{3+}$/Ir$^{5+}$ picture will accurately describe Sr$_2$Ir$_{1-x}$Rh$_x$O$_4$ at low dopings, and within the percolation regime that our RMXS measurements will focus on (i.e. for 0 $\le$ x $\le$ 0.24), at higher dopings this picture must be modified.  The complication arises from the fact that once the concentration of Rh reaches x = 0.50 there will no longer be enough potential Ir$^{5+}$ ions available to balance the electronic charge.  As a result, while the lower dopings will be dominated by Rh$^{3+}$ ions, the higher dopings must contain some mixture of 3+ and 4+ oxidation states.  This scenario appears to be consistent with the XAS fit parameters provided in Table I.  Although the position of the Rh L$_3$ edge white-line does not change between x = 0.07 and x = 0.70, the width of the white-line peak becomes significantly broader for x = 0.42 and x = 0.70.  This broadening is consistent with the development of a high-energy shoulder on the white-line peak, as one would expect for an increasing, but still minority, population of Rh$^{4+}$ ions. 

\begin{table}
\caption{ Doping dependence of Rh L$_3$-edge XAS parameters for Sr$_2$Ir$_{1-x}$Rh$_x$O$_4$.  The position and the width (FWHM) of the Rh L$_3$-edge white-line feature have been obtained from fits performed using a simple two component (Lorentzian + Arctangent) fit function.}

\begin{tabular}{c c c}
    \hline 
    \hline
    Rh Concentration & Peak Position (eV) & Peak Width (eV) \\
    \hline
		x = 0.07 & 3005.8 $\pm$ 0.1 & 3.2 $\pm$ 0.1 \\
    x = 0.11 & 3005.8 $\pm$ 0.1 & 3.1 $\pm$ 0.1 \\ 
    x = 0.15 & 3005.8 $\pm$ 0.1 & 3.3 $\pm$ 0.1 \\ 
    x = 0.24 & 3005.9 $\pm$ 0.1 & 3.4 $\pm$ 0.1 \\
		x = 0.42 & 3005.9 $\pm$ 0.1 & 4.1 $\pm$ 0.1 \\ 
    x = 0.70 & 3005.9 $\pm$ 0.1 & 3.9 $\pm$ 0.1 \\
		x = 1.00 & 3007.2 $\pm$ 0.1 & 3.9 $\pm$ 0.2 \\
    \hline
    \hline
\end{tabular}
\end{table}

\subsection{Magnetic Structure of S\lowercase{r}$_2$I\lowercase{r}$_{1-x}$R\lowercase{h}$_{x}$O$_4$}

The impact of Rh-doping on magnetic structure was investigated using resonant magnetic x-ray scattering (RMXS).  The magnetic ground state of pure Sr$_2$IrO$_4$ is known to be a canted {\it ab}-plane antiferromagnet in which magnetic moments follow the rotations of IrO$_6$ octahedra \cite{Kim_Science_2009,Ye_PRB_2013}.  This structure, AF-I, is illustrated in Fig. 2(a).  The magnetic structure of Sr$_2$Ir$_{1-x}$Rh$_x$O$_4$ was determined using three different elements from the RMXS data: (1) the magnetic selection rule, (2) the magnetic structure factor, and (3) the azimuthal dependence of the magnetic Bragg peaks.

\begin{figure}
\includegraphics{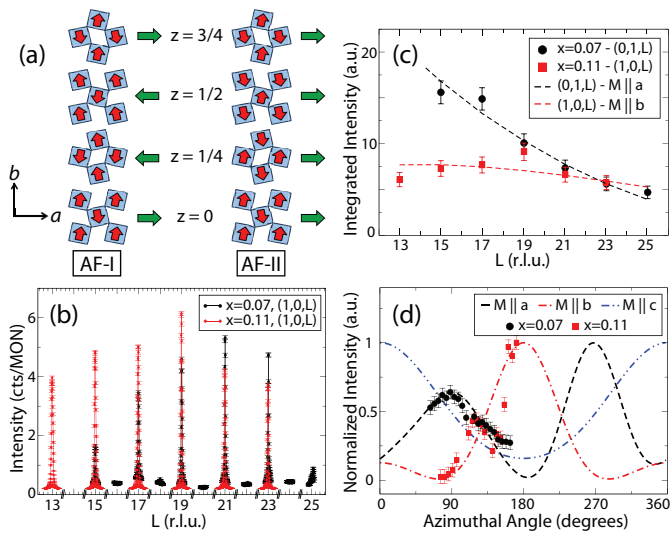}
\caption{(Color Online) (a) Canted antiferromagnetic ground states of Sr$_2$Ir$_{1-x}$Rh$_x$O$_4$.  The AF-I spin configuration is observed for x = 0, while the AF-II configuration is observed for 0.07 $\le$ x $\le$ 0.15.  The net ferromagnetic moment in each Ir-O layer is denoted by a green arrow.  (b) Observed magnetic Bragg peaks in Sr$_2$Ir$_{1-x}$Rh$_x$O$_4$ for x = 0.07 and x = 0.11. (c) Theoretical modelling of the magnetic structure factor is consistent with a canted antiferromagnetic structure (AF-II) which has a magnetic easy-axis in the {\it ab}-plane. (d) The azimuthal dependence of the (0,1,21) magnetic peak intensity confirms that the orientation of the moments is along the {\it a} and {\it b}-axes.  The data in panels (b)-(d) was collected at T = 7 K and E$_i$ = 11.217 keV, using $\sigma$-$\pi$ polarization analysis.}
\end{figure}

The simplest of these elements is the magnetic selection rule, which is illustrated in Fig. 2(b).  This panel shows the characteristic magnetic Bragg peaks which develop in Sr$_2$Ir$_{1-x}$Rh$_x$O$_4$ for dopings of x = 0.07 and 0.11.  Note that on the basis of this selection rule alone, one can immediately identify that a doping-induced magnetic phase transition takes place between x = 0 and x = 0.07.  The AF-I magnetic structure which develops in undoped Sr$_2$IrO$_4$ (x = 0) gives rise to magnetic Bragg peaks at (1,0,L)/(0,1,L) wave vectors for L = even, and (0,0,L) wave vectors for L = odd.  A single AF-I domain produces magnetic peaks at (1,0,4n+2) and (0,1,4n) wave vectors for all integer n.  However, given the tetragonal crystal structure of Sr$_2$IrO$_4$, it is natural for a two-domain magnetic structure to develop, with the second domain giving rise to peaks at (1,0,4n) and (0,1,4n+2).  This selection rule is clearly inconsistent with the data in Fig. 2(b), allowing us to rule out the possibility of an AF-I spin configuration for x = 0.07 and x = 0.11.

The magnetic Bragg peaks observed in Sr$_2$Ir$_{1-x}$Rh$_{x}$O$_{4}$ (0.07 $\le$ x $\le$ 0.15) appear at (1,0,L)/(0,1,L) wave vectors for L = odd.  Scans along other high symmetry directions in reciprocal space, such as [0,0,L], [1,1,L], [1/2,1/2,L], and [1/2,0,L], reveal no evidence of additional magnetic peaks, either at commensurate or incommensurate wave vectors.  These magnetic peaks are consistent with a {\bf k} = (0,0,0) magnetic propagation vector.  Following a similar approach to Calder et al \cite{Calder_PRB_2012}, we used representational analysis to identify potential magnetic structures for Sr$_2$Ir$_{1-x}$Rh$_x$O$_4$.  This analysis was performed using the SARAh Representational Analysis software package \cite{SARAh}.  For a crystal structure with {\it I4$_1$/acd} symmetry and magnetic moments located on the Ir 8a site, there are only six irreducible representations consistent with a propagation vector of {\bf k} = (0,0,0): $\Gamma_1$, $\Gamma_3$, $\Gamma_6$, $\Gamma_8$, $\Gamma_9$, and $\Gamma_{10}$.  Two of these representations can immediately be discarded as they fail to reproduce the observed magnetic Bragg peaks - $\Gamma_3$ (which describes a magnetic structure with ferromagnetic in-plane coupling, ferromagnetic out-of-plane coupling, and moments oriented along the {\it c}-axis), and $\Gamma_6$ (which describes a magnetic structure with ferromagnetic in-plane coupling, antiferromagnetic out-of-plane coupling, and moments oriented along the {\it c}-axis).  The four remaining irreducible representations ($\Gamma_1$, $\Gamma_8$, $\Gamma_9$, and $\Gamma_{10}$) are all characterized by antiferromagnetic in-plane coupling, and antiferromagnetic out-of-plane coupling.  The chief distinction between these representations is the choice of magnetic easy axis.  $\Gamma_1$ and $\Gamma_8$ describe magnetic structures with moments oriented along the {\it c}-axis (as in the doping-induced state observed in Sr$_2$Ir$_{0.9}$Mn$_{0.1}$O$_4$ [Ref. 14], while $\Gamma_9$ and $\Gamma_{10}$ describe structures with moments in the {\it ab}-plane (as in the field-induced state of Sr$_2$IrO$_4$ [Ref. 2]).  The magnetic structure corresponding to $\Gamma_9$ and $\Gamma_{10}$, which we will label AF-II, is illustrated in Fig. 2(a). 

In order to distinguish between these possible structures, we can model both the magnetic structure factor and the azimuthal dependence of the magnetic Bragg peaks.  These quantities are both sensitive to the orientation of the magnetic moments, and can be calculated using the FDMNES software package \cite{FDMNES}.  For simplicity, we have employed a single-domain model for these calculations, which assumes one dominant magnetic domain.  The integrated intensity of the magnetic Bragg peaks (obtained from $\theta$ or rocking scans) is plotted as a function of L in Fig. 2(c).  These measurements were performed with the sample aligned such that the [1,1,0] and [0,0,1] directions are coincident with the vertical scattering plane.  In this orientation, there will be non-zero magnetic scattering contributions from moments aligned along the {\it a}, {\it b}, or {\it c}-axes.  However, a satisfactory fit to the magnetic structure factor can only be obtained for moments oriented within the {\it ab}-plane - either along the {\it a}-axis (x = 0.07) or along the {\it b}-axis (x = 0.11).  Note that because the crystal structure of Sr$_2$Ir$_{1-x}$Rh$_x$O$_4$ has tetragonal symmetry, there is no physical distinction between these two axes.  Hence the apparent 90$^{\circ}$ rotation of moment direction between x = 0.07 and x = 0.11 is simply due to a spontaneous choice of [1,0,0]/[0,1,0] orientation adopted by the dominant grain upon cooling through T$_{N1}$.

The azimuthal dependence of the magnetic Bragg peaks in Sr$_2$Ir$_{1-x}$Rh$_x$O$_4$ (Fig. 2(d)) is also consistent with magnetic moments oriented in the {\it ab}-plane.  Here $\phi$ = 0$^{\circ}$ has been defined as the sample orientation for which [0,1,0] is coincident with the vertical scattering plane defined by {\bf k$_i$} and {\bf k$_f$}.  The modeling of the azimuthal dependence indicates that the magnetic easy axis is along the {\it a}-axis for x = 0.07, and along the {\it b}-axis for x = 0.11, in full agreement with the results of the structure factor calculation.  In particular, two qualitative features of the azimuthal dependence - the 180$^{\circ}$ oscillation period and the vanishing of magnetic intensity at specific angles - cannot be reproduced by a magnetic structure which has a {\it c}-axis spin configuration.

The results of our magnetic structure analysis indicate that: (1) Sr$_2$Ir$_x$Rh$_{1-x}$O$_4$ undergoes a doping-induced magnetic phase transition at x $\le$ 0.07, (2) the magnetic ground state of Sr$_2$IrO$_4$ is very sensitive to a variety of external perturbations, and (3) the effects of quenched magnetic (Mn) and non-magnetic (Rh) impurities are significantly different.  It should be noted that a full magnetic structure factor and azimuthal dependence measurement was not completed for the x = 0.15 sample.  We have attributed the same AF-II magnetic structure to this compound based purely on the magnetic selection rule.  As in the case of the x = 0.07 and x = 0.11 compounds, this sample displays magnetic Bragg peaks at (1,0,L) and (0,1,L) wave vectors for L = odd, but not for L = even.  Additional follow-up measurements would be required for an unambiguous determination of the magnetic structure and moment direction for this doping.

\subsection{Magnetic Order Parameter and Correlation Lengths in S\lowercase{r}$_2$I\lowercase{r}$_{1-x}$R\lowercase{h}$_{x}$O$_4$}

The temperature dependence of the magnetic peak intensity (Fig. 3) provides a direct measure of the antiferromagnetic order parameter (I $\sim$ M$^2$).  The magnetic peak intensity closely tracks the bulk magnetization \cite{Qi_PRB_2012}, with T$_{N1}$ marking the appearance of magnetic Bragg peaks and a net ferromagnetic moment, and T$_{N2}$ marking a dramatic increase in peak intensity and a magnetization kink.  Although the magnetic peaks persist between T$_{N1}$ and T$_{N2}$, they display a broadened lineshape which is indicative of finite magnetic correlation lengths.  By combining our RMXS measurements with previously reported magnetization data\cite{Qi_PRB_2012}, we can construct the magnetic phase diagram provided in Fig. 4(a).

\begin{figure}
\includegraphics{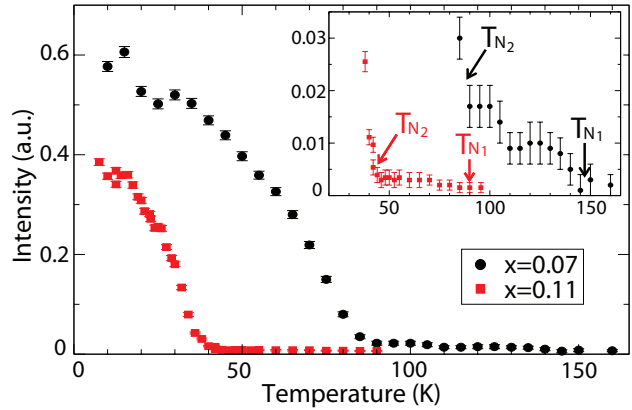}
\caption{(Color Online) Temperature dependence of the (0,1,21) magnetic Bragg peak measured at the Ir L$_3$-edge.  The magnetic peak intensity is significantly reduced above T$_{N2}$, but remains finite up to T$_{N1}$ (as shown in the inset).  Magnetic peak intensities from the x = 0.07 and x = 0.11 samples have been normalized with respect to each other, using the intensities of nearby structural Bragg peaks for reference.  Data in this panel was collected at E$_i$ = 11.217 keV, using $\sigma$-$\pi$ polarization analysis.}
\end{figure}

\begin{figure}
\includegraphics{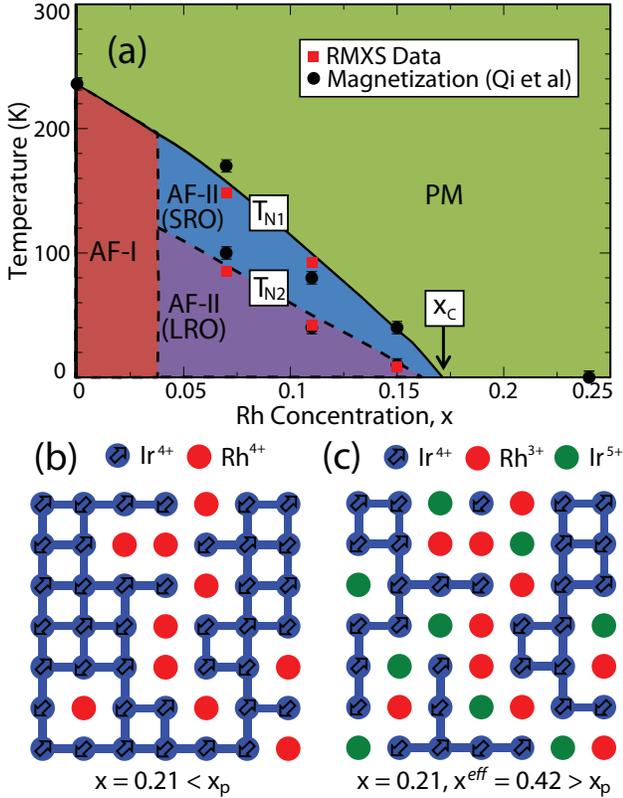}
\caption{(Color Online) (a) The magnetic phase diagram of Sr$_2$Ir$_{1-x}$Rh$_x$O$_4$, as constructed from RMXS and previously reported magnetization results \cite{Qi_PRB_2012}.  The disappearance of magnetic order at x$_c$ $\sim$ 0.17 can be understood in terms of a spin-orbital percolation picture.  (b)  If Ir$^{4+}$ is substituted for Rh$^{4+}$, then each dopant ion introduces one magnetic/orbital defect (replacing a $J_{eff}$ = 1/2 Ir moment with a S = 1/2 Rh moment).  In this scenario, a doping of x = 0.21 falls well below the spin-only percolation threshold of x$_p$ = 0.407. (c)  Alternatively, if Ir$^{4+}$ is substituted for Rh$^{3+}$, then each dopant ion introduces two magnetic/orbital defects (one from the Rh$^{3+}$ [S = 0], and one from its Ir$^{5+}$ [$J_{eff}$ = 0] ionic counterpart).  In this scenario, the same doping (x = 0.21) is now sufficient to exceed the percolation threshold and destroy magnetic order.  The discrepancy between 2x$_c$ and x$_p$ reflects the importance of orbital percolation effects, which arise due to the strong SOC of this system.}
\end{figure}

These results suggest that the magnetic phase diagram of Sr$_2$Ir$_{1-x}$Rh$_{x}$O$_4$ is characterized by two distinct regions of AF-II magnetic order; a long-range-ordered (LRO) phase below T$_{N2}$, and a short-range-ordered (SRO) phase between T$_{N1}$ and T$_{N2}$.  The change in magnetic correlation lengths at T$_{N2}$ is reflected in the width of the magnetic Bragg peaks, as shown in Fig. 5.  The magnetic peaks within the SRO phase are significantly weaker than those observed in the LRO phase, and appear to be broader along both the in-plane ([H,0,0] and [0,K,0]) and out-of-plane ([0,0,L]) directions.  The experimentally measured peak width (expressed as the FWHM, $\Gamma_{obs}$) represents a convolution of the intrinsic peak width ($\Gamma_{int}$) and the instrumental resolution function ($\Gamma_{res}$).  In this case, an experimental resolution function was determined by measuring the lineshape of a nearby structural Bragg peak.  For the L-scans provided in Fig.5, the width of the experimental resolution function was $\Gamma_{res}$ $\sim$ 0.0066 r.l.u.  The magnetic correlation length ($\xi$) is inversely proportional to the intrinsic peak width through the relation: $\xi$ = [(2$\pi$/d)($\Gamma_{int}$/2)]$^{-1}$.  This allows us to determine the average magnetic correlation lengths within the SRO phase, which are found to be $\xi_{ab}$ $\sim$ 1500 {\AA} (x = 0.07) and 1400 {\AA} (x = 0.11) in-plane, and $\xi_{c}$ $\sim$ 1000 {\AA} (x = 0.07) and 800 {\AA} (x = 0.11) out-of-plane.  Note that in both of these samples the average magnetic correlation length is substantially longer than the average distance between Rh dopant ions ($\sim$ 95 {\AA} and 60 {\AA}, respectively).

\begin{figure}
\includegraphics{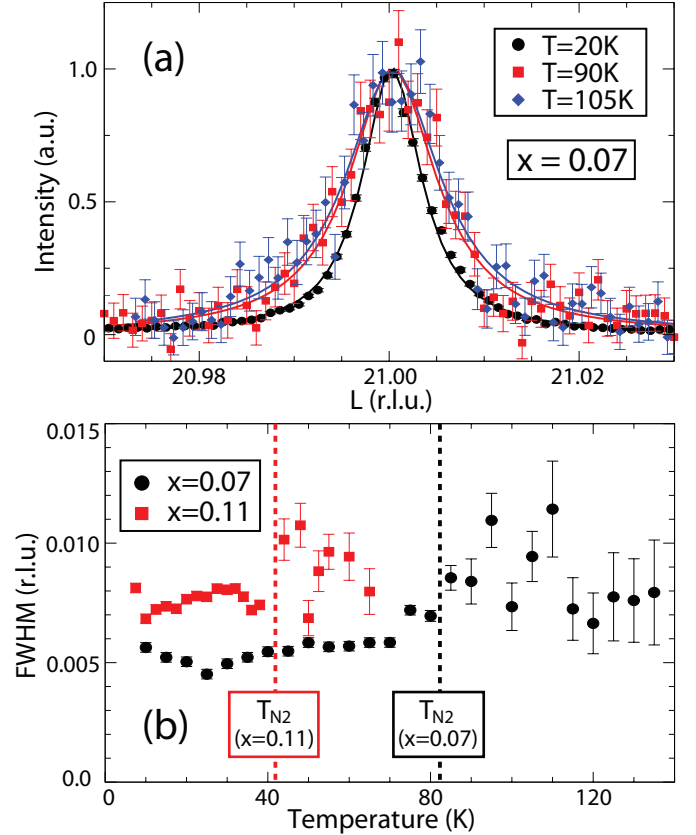}
\caption{ (Color Online) Magnetic correlation lengths in Sr$_2$Ir$_{1-x}$Rh$_x$O$_4$.  (a) Reciprocal space scans through the (0,1,21) magnetic Bragg peak in Sr$_2$Ir$_{0.93}$Rh$_{0.07}$O$_4$ at temperatures below (T = 20 K) and above (T = 90, 105 K) the magnetic transition at T$_{N2}$.  (b) Temperature dependence of the magnetic peak width along [0,0,L] for samples with x = 0.07 and x = 0.11.  The FWHM in this panel represents the intrinsic peak width, determined from resolution-convoluted fits.  The broadening of the magnetic peak widths above T$_{N2}$ indicates the presence of finite magnetic correlation lengths between T$_{N1}$ and T$_{N2}$.  The correlation lengths within this phase range from 800 to 1000 {\AA}. }
\end{figure}

To summarize, the magnetic phase diagram of Sr$_2$Ir$_{1-x}$Rh$_x$O$_4$ is distinguished by three major features: (1) the disappearance of magnetic order at a critical doping of x$_c$ $\sim$ 0.17, (2) a doping-induced change in magnetic structure between x = 0 and x = 0.07, and (3) a thermally-driven transition between long-range (LRO) and short-range (SRO) magnetic order at T$_{N2}$.

\subsection{Robustness of the $J_{eff}$ = 1/2 Ground State}

The RMXS data also allows us to address the question of how Rh-doping affects the $J_{eff}$ = 1/2 character of Sr$_2$IrO$_4$.  In previous work \cite{Kim_Science_2009,Boseggia_PRB_2012,JWKim_PRL_2012,Calder_PRB_2012,Boseggia_PRL_2013,Ohgushi_PRL_2013}, the $J_{eff}$ = 1/2 ground state has been identified on the basis of an anomalously large L$_3$/L$_2$ magnetic intensity ratio, which arises due to the selection rules and transition matrix elements associated with the L$_2$ (2p$_{1/2}$ $\rightarrow$ 5d$_{3/2}$) and L$_3$ (2p$_{3/2}$ $\rightarrow$ 5d$_{3/2}$, 5d$_{5/2}$) RMXS processes.  The energy dependence of the (0,1,21) magnetic Bragg peak in Sr$_2$Ir$_{1-x}$Rh$_x$O$_4$ is provided in Fig. 6.  Note that extremely large L$_3$/L$_2$ intensity ratios are observed for both the x = 0.07 and x = 0.11 samples.  In fact, no magnetic Bragg peaks could be detected at the L$_2$ edge for either sample, indicating that I(L$_3$)/I(L$_2$) $>$ 200.  A similar result has also been reported for Sr$_2$Ir$_{0.9}$Mn$_{0.1}$O$_4$ [Ref. 14], suggesting that the $J_{eff}$ = 1/2 character of Sr$_2$IrO$_4$ is very robust against doping in general.  This persistence of strong $J_{eff}$ character implies that the electronic transition at x $\sim$ 0.16 is not driven by the tuning of SOC effects, but rather by a combination of hole-doping and/or doping-induced structural changes \cite{Qi_PRB_2012}.  

It should be noted that the interpretation of the L$_3$/L$_2$ magnetic intensity ratio has recently been questioned by Chapon and Lovesey \cite{Chapon_JPCM_2011} and Moretti Sala et al \cite{Moretti_arXiv_2013}.  In particular, it has been suggested that the magnetic intensity at the L$_2$-edge may vanish if Ir$^{4+}$ magnetic moments are aligned within the {\it ab}-plane, regardless of the splitting of the t$_{2g}$ levels.  This point is relevant to both Sr$_2$IrO$_4$ and Sr$_2$Ir$_{1-x}$Rh$_x$O$_4$, as both systems adopt canted {\it ab}-plane antiferromagnetic ground states below T$_N$.  However, in the case of Sr$_2$Ir$_{0.90}$Mn$_{0.10}$O$_4$, which displays a collinear c-axis antiferromagnetic structure \cite{Calder_PRB_2012}, this objection does not apply.  In addition, Mn-doping represents an even stronger pertubation to magnetism ($J_{eff}$ = 1/2 $\rightarrow$ S = 3/2) and SOC (5d $\rightarrow$ 3d) than Rh-doping.  Although the signatures of the $J_{eff}$ = 1/2 state in Sr$_2$Ir$_{1-x}$Rh$_x$O$_4$ may still require further investigation, the analogy with Sr$_2$Ir$_{1-x}$Mn$_{x}$O$_4$ suggests that, at least on a qualitative level, these conclusions will still hold true.

\begin{figure}
\includegraphics{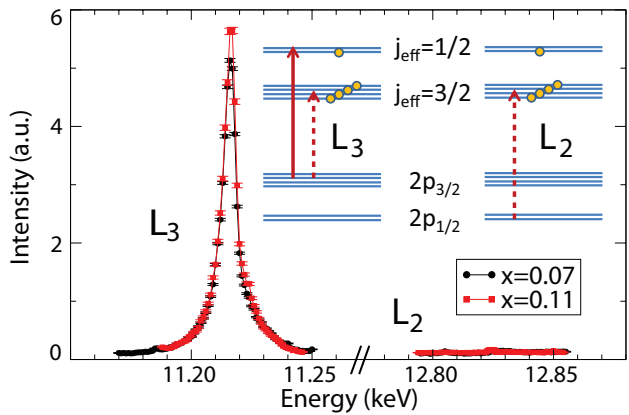}
\caption{(Color Online) Energy dependence of the (0,1,21) magnetic Bragg peak measured at the Ir L$_3$ (E = 11.215 keV) and L$_2$ (E = 12.824 keV) absorption edges.  The large L$_3$/L$_2$ intensity ratio associated with the $J_{eff}$ = 1/2 ground state remains unaffected by Rh concentrations up to x = 0.11.  All data presented in this figure was collected at T = 7 K. }
\end{figure}

\subsection{Octahedral Rotations and Structural Disorder in S\lowercase{r}$_2$I\lowercase{r}$_{1-x}$R\lowercase{h}$_{x}$O$_4$}

\begin{figure}
\includegraphics{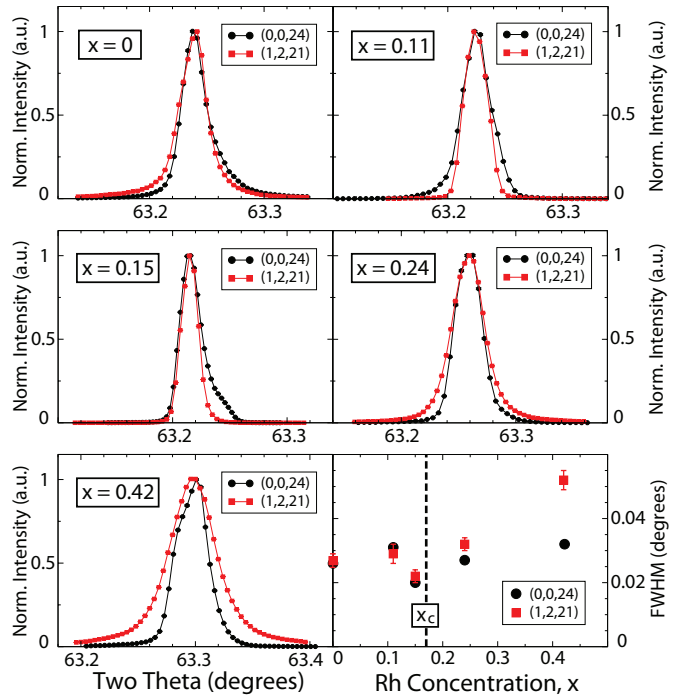}
\caption{ (Color Online) High-resolution x-ray diffraction measurements on Sr$_2$Ir$_{1-x}$Rh$_x$O$_4$.  This figure presents a collection of longitudinal ($\theta$-2$\theta$) scans through the (0,0,24) structural Bragg peak and (1,2,21) superlattice Bragg peak for samples with x = 0, 0.11, 0.15, 0.24, and 0.42.  For illustrative purposes, scans through the (0,0,24) peak (2$\theta$ $\sim$ 61.8$^{\circ}$) have been horizontally translated and centered.  The (1,2,21) superlattice peak arises due to correlated rotations of the IrO$_6$/RhO$_6$ octahedra.  The width of the superlattice peak is approximately equal to that of the Bragg peak for 0 $\le$ x $\le$ 0.24, indicating that the IrO$_6$/RhO$_6$ rotations are well-correlated at low dopings.  By x = 0.42 the superlattice peak is considerably broader, indicating significant rotational disorder at higher dopings.  All data presented in this figure was collected at T = 300 K.}
\end{figure}

The rotations of IrO$_6$ octahedra are known to play an important role in shaping the physics of Sr$_2$IrO$_4$.  These rotations break the inversion symmetry between nearest-neighbor Ir ions, giving rise to an antisymmetric Dzyaloshinskii-Moriya interaction \cite{Jackeli_PRL_2009}.  In addition, the orientation of the ordered moments in Sr$_2$IrO$_4$ appears to be strongly coupled to the octahedral rotations, with the canted antiferromagnetic ground state displaying a spin-canting angle of $\sim$ 8$^{\circ}$ [Refs. 2, 24].  Given the difference in rotation angles between Sr$_2$IrO$_4$ ($\sim$ 11$^{\circ}$) and Sr$_2$RhO$_4$ ($\sim$ 9.7$^{\circ}$), the disorder of IrO$_6$/RhO$_6$ octahedral rotations has been proposed as one possible explanation for the suppression of magnetic order in Sr$_2$Ir$_{1-x}$Rh$_x$O$_4$ [Ref. 9].
 
In order to investigate how the IrO$_6$/RhO$_6$ octahedral rotations in Sr$_2$Ir$_{1-x}$Rh$_x$O$_4$ evolve as a function of doping, we performed a series of high-resolution non-resonant x-ray diffraction measurements.  Fig. 7 provides a comparison of longitudinal ($\theta$-2$\theta$) scans taken through the (0,0,24) structural Bragg peak and the (1,2,21) superlattice Bragg peak for samples with x = 0, 0.11, 0.15, 0.24, and 0.42.  The (1,2,21) superlattice peak arises due to the correlated rotations of the IrO$_6$ octahedra, and it is one of the distinguishing characteristics of the {\it I4$_1$/acd} spacegroup \cite{Crawford_PRB_1994,Huang_JSSC_1994}.  In the absence of correlated octahedral rotations, the superlattice peaks at (1,2,L)/(2,1,L), L = odd, disappear and Sr$_2$IrO$_4$ can be described by an {\it I4/mmm} spacegroup, with a unit cell reduced by $\sqrt{2} \times \sqrt{2}$ in the {\it ab}-plane and halved along the {\it c}-axis.  As a result, the width of the (1,2,21) superlattice peak provides a window into the correlation lengths associated with these octahedral rotations.  For dopings of x = 0 to x = 0.24 the width of the superlattice peak is essentially the same as that of the structural Bragg peak, implying that the correlation lengths are long-ranged and the octahedral rotations are well-ordered.  At higher dopings (x = 0.42), the superlattice peak becomes significantly broader than the Bragg peak, indicating reduced octahedral correlation lengths ($\xi_{rot}$ $\sim$ 500 {\AA}) and increased rotational disorder.  This suggests that while rotational disorder may play an important role in Sr$_2$Ir$_{1-x}$Rh$_x$O$_4$ at higher dopings (x $\ge$ 0.42), it is not a significant effect at lower dopings (x $\le$ 0.24), and is unlikely to drive the suppression of magnetic order at x$_c$.

\section{Discussion and Conclusions}

It is very interesting to consider the mechanism responsible for the disappearance of magnetic order at x$_{c}$ $\sim$ 0.17.  We have already touched upon two potential mechanisms for this transition in the preceding sections.  From the lack of doping-dependence associated with the L$_3$/L$_2$ magnetic intensity ratio (Section III.D), we infer that this magnetic transition is not the result of spin-orbit tuning.  Similarly, although it is possible for magnetic order to be disrupted by the disorder of IrO$_6$/RhO$_6$ octahedral rotations \cite{Qi_PRB_2012}, our measurements of the superlattice Bragg peaks associated with these rotations (Section III.E) reveal no significant change in correlation lengths at x$_c$.  Other doping-induced structural changes, such as a sudden jump in Ir-O-Ir bond angle, have previously been reported in the vicinity of x$_c$ [Ref. 9].  However, these structural changes appear to be discontinuous, while the observed decrease in T$_{N1}$ and T$_{N2}$ is clearly continuous.

An alternative explanation is provided by percolation theory (Figs. 4(b,c)), which has proven extremely successful at describing magnetism in doped cuprates such as La$_2$Cu$_{1-x}$(Zn,Mg)$_x$O$_4$ [Ref. 38].  This argument assumes that Sr$_2$Ir$_{1-x}$Rh$_x$O$_4$ can be adequately described by a local moment picture for x $\le$ 0.24, a claim which appears reasonably well-justified based on previous resistivity data \cite{Qi_PRB_2012}.  As Sr$_2$IrO$_4$ is effectively a two-dimensional S = 1/2 Heisenberg antiferromagnet, we expect the conventional (i.e. spin-only) percolation threshold for this system to be x$_p$ = 0.407 [Ref. 39].  Since each Rh$^{3+}$ dopant ion introduces two non-magnetic vacancies, the effective site dilution, x$^{eff}$, will be equal to twice the nominal Rh concentration.  We suggest that the apparent discrepancy between x$^{eff}_c$ = 2x$_c$ = 0.34 and x$_p$ = 0.407 may reflect novel percolation behavior arising from strong SOC effects.  Recent theoretical work \cite{Tanaka_PRL_2007,Tanaka_PRB_2009} has shown that quantum orbital systems are much more sensitive to site dilution than pure spin systems, and can display considerably lower percolation thresholds.  In a system where the spin and orbital degrees of freedom are strongly entangled, as in Sr$_2$IrO$_4$, it is therefore unsurprising that a spin-only percolation calculation overestimates the value of x$_p$.  This result suggests a full theoretical description of dilute magnetism in Sr$_2$Ir$_{1-x}$Rh$_x$O$_4$ must account for both spin and orbital percolation effects, raising the possibility of exciting new percolation physics in the strong SOC regime.

In conclusion, we have used a combination of resonant x-ray techniques to investigate the chemical, electronic, and magnetic properties of the doped spin-orbital Mott insulator Sr$_2$Ir$_{1-x}$Rh$_x$O$_4$.  XAS measurements clearly demonstrate that Rh-doping introduces Rh$^{3+}$ and Ir$^{5+}$ ions into this material, leading to (1) hole-doping and (2) magnetic dilution of the system.  RMXS measurements reveal a doping-induced change in magnetic structure at x $\le$ 0.07, which leads to the development of a canted {\it ab}-plane antiferromagnetic state (AF-II) for x = 0.07, 0.11, and 0.15.  Magnetic order is fully suppressed above x$_c$ $\sim$ 0.17 (or x$^{eff}_c$ $\sim$ 0.34), a result which suggests novel percolation effects and intriguing differences from diluted cuprates.  We hope these results will help to motivate further theoretical and experimental work on Sr$_2$Ir$_{1-x}$Rh$_x$O$_4$ and other doped 5d systems in the future.

\begin{acknowledgments}
The authors would like to acknowledge valuable discussions with Y. Cao, D. Dessau, D. Haskel, and J.W. Kim.  Work at the University of Toronto was supported by NSERC of Canada, the Banting Postdoctoral Fellowship program, and the Canada Research Chair program.  Work at the University of Kentucky was supported by NSF through grants DMR-0856234, DMR-1265162, and EPS-0814194.  Use of SXRMB at the Canadian Light Source is supported by NSERC of Canada, NRC of Canada, CIHR, and the University of Saskatchewan.  Use of the Advanced Photon Source at Argonne National Laboratory and the National Synchrotron Light Source at Brookhaven National Laboratory is supported by the U.S. Department of Energy, Office of Science, Office of Basic Energy Sciences, under Contract Nos. DE-AC02-06CH11357 and DE-AC02-98CH10886.
\end{acknowledgments}

\end{document}